\documentclass[a4paper]{jpconf}
\usepackage{graphicx}
\begin{document}
\title{The NA61/SHINE long target pilot analysis for T2K}

\author{Nicolas Abgrall\\
on behalf on the NA61/SHINE collaboration}

\address{D\'epartement de Physique Nucl\'eaire et Corpusculaire
  (DPNC), University of Geneva, 24 quai Ernest Ansermet, 1205, Geneva, Switzerland}

\ead{nicolas.abgrall@cern.ch}

\begin{abstract}
  The NA61/SHINE collaboration performed measurements of pC
  interactions at 31~GeV/c beam momentum with a full size replica of
  the T2K target (1.9 interaction length) during a pilot run in
  2007. Larger statistics runs were also conducted in 2009 and 2010.
  The NA61/SHINE setup consists in a large acceptance spectrometer
  located on the H2 beamline of the SPS at CERN. For the first time,
  the kinematical phase space of interest for an
  accelerator based neutrino experiment (i.e. kinematical phase space of
  pions/kaons exiting the target and producing neutrinos in the
  direction of the near and far detectors) is fully covered by a single hadron
  production experiment. In a first stage, yields of positively
  charged pions were measured at the surface of the target. The
  analysis of the 2007 data set presented here demonstrates that {\bf
    a)} high quality long target data were successfully taken with the
  NA61/SHINE apparatus, and {\bf b)} for the first time, the T2K
  neutrino flux predictions can effectively be re-weighted with the
  NA61/SHINE long target data.\\
{\it Invited paper to NUFACT11, XIIIth International Workshop on
  Neutrino Factories, Super Beams and Beta Beams, 1-6 Aug. 2011, CERN
  and University of Geneva, submitted to IOP conference series.}
\end{abstract}
\vspace{-2pc}
\section{Motivations for long target measurements for T2K}
A prediction of the neutrino beam content at the T2K (\cite{T2K-exp}, \cite{T2K-NIM}) far detector is
given in Table \ref{beam-content-far}.  The $\nu_\mu$ flux is
predominantly ($\sim 95$\%) produced by the decay in flight of
positively charged pions focused by the magnetic horns of the
beamline. While $\sim 30$\% of the $\nu_e$ flux is produced from the
decay of positively charged kaons, $\sim 50$\% of it is
due to muons produced in the decay of the same charged pions that
generate the $\nu_\mu$ flux. Thus, in a first stage, pion production
data will constrain both $\nu_\mu$ and $\nu_e$ fluxes.
\vspace{-1.5pc}
\begin{table}[h]
  \centering
  \caption{\label{beam-content-far} Composition of the neutrino beam and its various species at the far
    detector. Integrated values are quoted. Predictions obtained with the
    GCALOR model and horn currents set to 320 kA each (T2K was running
    with horn currents of 250 kA in 2010/11).}
\vspace{0.2pc}
\scalebox{0.90}{
	\begin{tabular}{ccccccccccccc}
	\hline
        &   &  &\multicolumn{8}{c}{Source}\\ \cline{4-13}
	\multicolumn{1}{c}{$\nu$} & \multicolumn{2}{c}{Flux} &
        \multicolumn{2}{c}{$\pi^{+}$ or $\pi^{-}$} 
        & \multicolumn{2}{c}{K$^{+}$ or K$^{-}$ (K2)} 
        & \multicolumn{2}{c}{K$^{+}$ or K$^{-}$ (K3)} &
        \multicolumn{2}{c}{K$^{0}_{L}$} 
        & \multicolumn{2}{c}{$\mu^{+}$ or $\mu^{-}$}\\ \cline{2-13}
	species & Abund. & $\langle E_{\nu}\rangle$ & $\%$ & $\langle E_{\nu}\rangle$  &
        $\%$ & $\langle E_{\nu}\rangle$ & $\%$ & $\langle
        E_{\nu}\rangle$ & $\%$ & $\langle E_{\nu}\rangle$ & $\%$ & $\langle E_{\nu}\rangle$ \\
        \hline
        $\nu_{\mu}$       & 1.0    & 0.79 & 95.1 & 0.64 & 4.5  & 4.0
        & 0.24 & 1.93 & 0.1 & 2.05 & 0.01 & 0.75\\
        $\bar \nu_{\mu}$ & 0.0701  & 1.14 & 85.8 & 1.05 & 4.6  &
        3.1 & 0.2   & 1.56    & 1.3     & 2.05 & 8.0 & 0.68\\
        $\nu_{e}$         & 0.0110 & 1.40 &  1.0  &  1.48 & -- & --  & 33.0 &
        2.25 & 12.5  & 2.38 & 53.3 & 0.64 \\
        $\bar \nu_{e}$   & 0.0017 & 2.18 &  0.4 &  2.32 & -- & -- & 14.7 & 
        1.84 & 77.6 & 2.38 & 7.2  & 0.75\\
        \hline
        \end{tabular}
}
\end{table}

\begin{figure}[h]
\label{motivations}
\includegraphics[width=12pc]{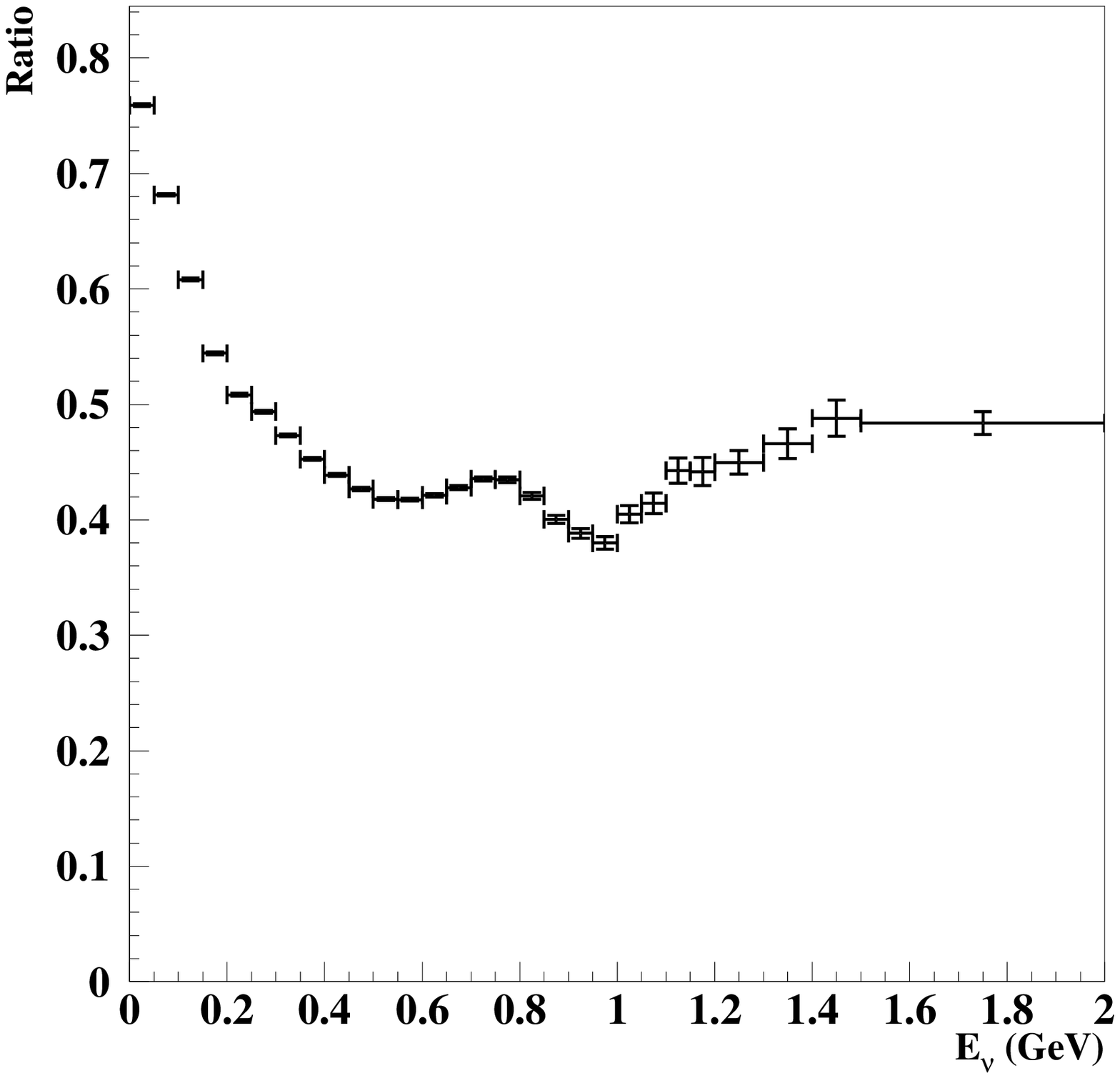}
\hspace{-1pc}
\includegraphics[width=12pc]{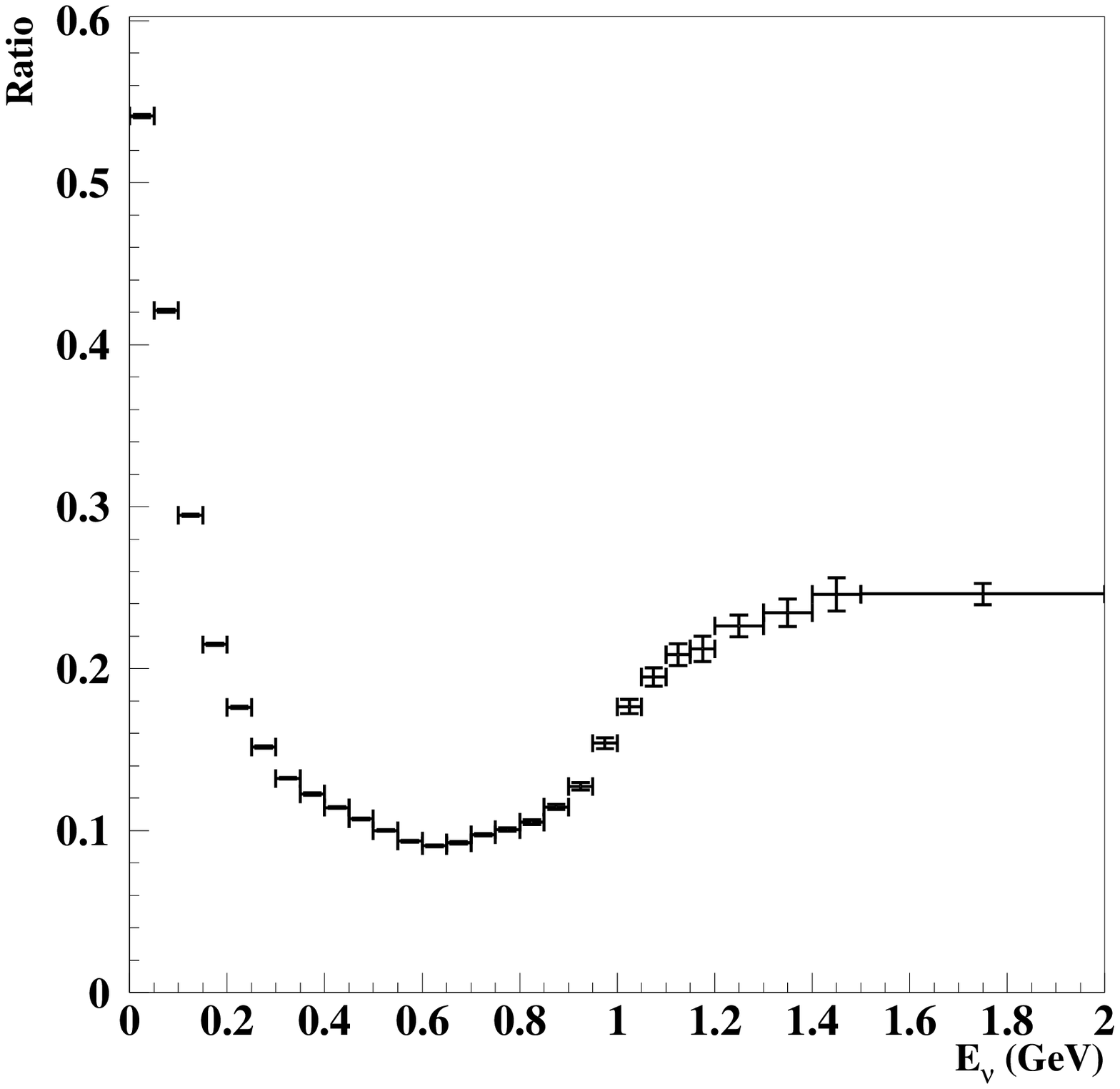}
\hspace{-1pc}
\includegraphics[width=14pc]{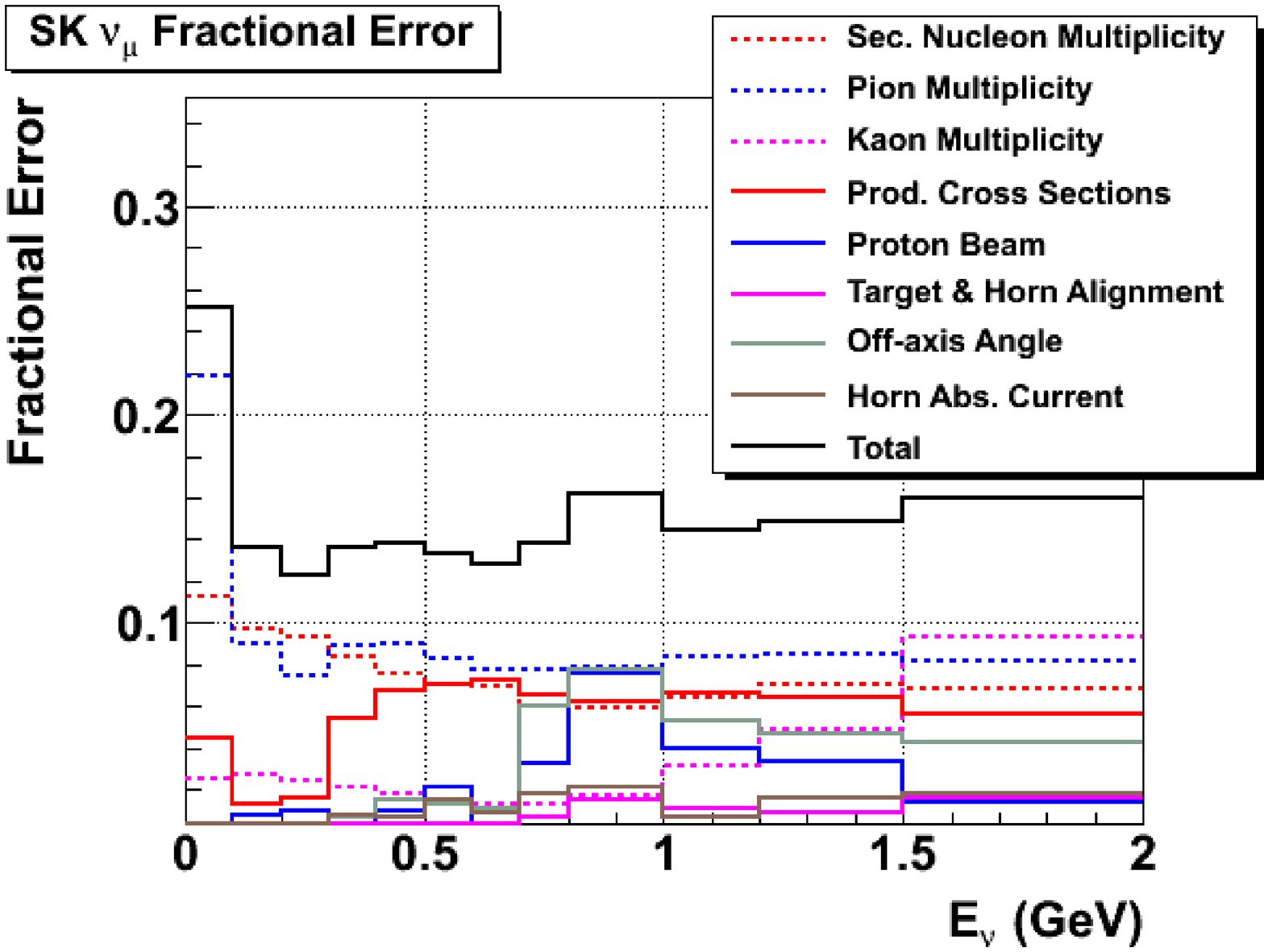}
\caption{Ratio of (1-direct) [left] and (1-target) [middle] to total
  contribution for the $\nu_\mu$ flux at the far detector.
  Current error envelopes for the $\nu_\mu$ flux at the far detector
  [right] (for the analysis described in \cite{T2K-nue-paper}).}
\end{figure}

In terms of hadron production measurements, the pion contribution to
the $\nu_\mu$ flux at the far detector can be decomposed into a {\it
  direct} component and a {\it target} component. The first
contribution originates from pions directly produced in the proton
primary interaction (secondary pions) or in the decay of a secondary
particle.  The {\it target} contribution refers to all pions exiting
the target or pions produced in the decay of an outgoing particle. The
dependence of these contributions on the neutrino energy is depicted
in Fig.~\ref{motivations} (left and middle panels). While the direct component
contributes to $\sim 60$\% of the $\nu_\mu$ flux at the beam peak
energy of $\sim 600$~MeV, the target component contributes to $\sim 90$\%
of the flux. Thus, thin target measurements ($p+C\rightarrow \pi^+ +
X$ at 30 GeV) for T2K constrain up to 60\% of the $\nu_\mu$ flux
prediction at the far detector, the remaining 40\% being produced in
secondary interactions in the target and elements of the
beamline. Long target measurements would constrain up to 90\% of the
flux prediction. In this case, only 10\% of the flux due to secondary interactions out of the
target would require constraints from other data.  \vspace{-2pc}

As depicted in Fig.~\ref{motivations} (right panel),
the total fractional error on the T2K $\nu_\mu$ flux
prediction (\cite{T2K-beam}, \cite{T2K-nue-paper}) at the far detector is $\sim 15$\% at the beam peak
energy. Uncertainties arise from two main contributions:
uncertainties related to the beamline setting (e.g. beam optics, beam direction,
alignment and currents of the focusing horns) and those related to the
re-weighting procedure based on the NA61/SHINE thin target
data (within 5 to 10~\%)\cite{NA61-pion-paper}. The fractional error at the beam peak
energy is dominated by uncertainties from the second contribution
(i.e. pion/kaon multiplicity, secondary nucleon multiplicity and
production cross sections). In particular, uncertainties on the
re-weighting of secondary nucleon and production cross sections for
secondary interactions are applied to $\sim 40$\%
($\sim 10$\%) of
the flux in the case of the thin (long) target based re-weighting. These
uncertainties rely on interpolation between sparse data sets or comparison to models and are
often poorly known. In the case where flux predictions are re-weighted by long target data,
these contributions are taken into
account at once and there is no need for error prone estimates
of secondary interactions in the target. Long target data are therefore needed to
obtain a precision of 5\% or better on the absolute flux prediction.

\section{The NA61/SHINE long target measurements for T2K}
The 2007 NA61/SHINE pilot data were taken with a replica of the T2K
target at 30 GeV beam energy.  The NA61/SHINE spectrometer consists in
a set of 5 time projection chambers (TPCs) complemented by an array of
time-of-flight (ToF) detectors located downstream of the TPCs,
allowing for a full coverage of the forward pion production of
interest for T2K (i.e. phase space of pions exiting the target and
producing neutrinos in the direction of the near and far detectors). Details of
the experimental setup are given elsewhere \cite{NA61-pion-paper}. The
trajectory of each beam proton track is reconstructed with a set of beam
position detectors that allow to determine the position of the beam
impact on the upstream face of the target. The usage of the ToF in the
analysis assures that selected tracks in the TPCs originate from
single proton interactions in the target. Data can therefore be
normalized to the total number of protons on target (POT).

Tracks reconstructed in the TPCs are extrapolated back through the
magnetic field from their first measured point to the surface of the
target. A point-of-closest-approach is found between the trajectory of
the track and the surface of the target. This requires a precise knowledge of the
relative alignment of
the beam and target, which is determined by using both beam and TPC tracks.
For the 2007 pilot run, the target was tilted with respect to the beam
axis in both horizontal ($\sim 5$ mrad) and vertical ($\sim 2.8$ mrad)
directions. The misalignment was taken into account in the extrapolation
procedure and effects on the outgoing pion yields (additional systematic uncertainty) studied with
dedicated Monte-Carlo simulations. The target is split into 5
bins of 18 cm each along the beam (z axis), and a last bin corresponding to the
dowstream face of the target that covers the very forward (below 40 mrad) pion
production.  Analysis cuts are optimized to improve the momentum and
polar angle resolutions at the first measured point on track that
determine the achievable resolutions on target, estimated to $\sigma_z=5$ cm,
$\sigma_p/p=3$\% and $\sigma_\theta/\theta=6$\% for the longitudinal,
momentum and polar angle resolutions respectively. As depicted in
Fig.~\ref{reconstruction} (left panel), tracks can effectively be
reconstructed on the surface of the target. The analysis coverage
extends up to 20 GeV/c in the forward production region ($<40$ mrad)
and covers angles up to 280 mrad at lower momentum.

\begin{figure}[h]
\label{reconstruction}
\vspace{-2pc}
\includegraphics[width=13pc,height=12pc]{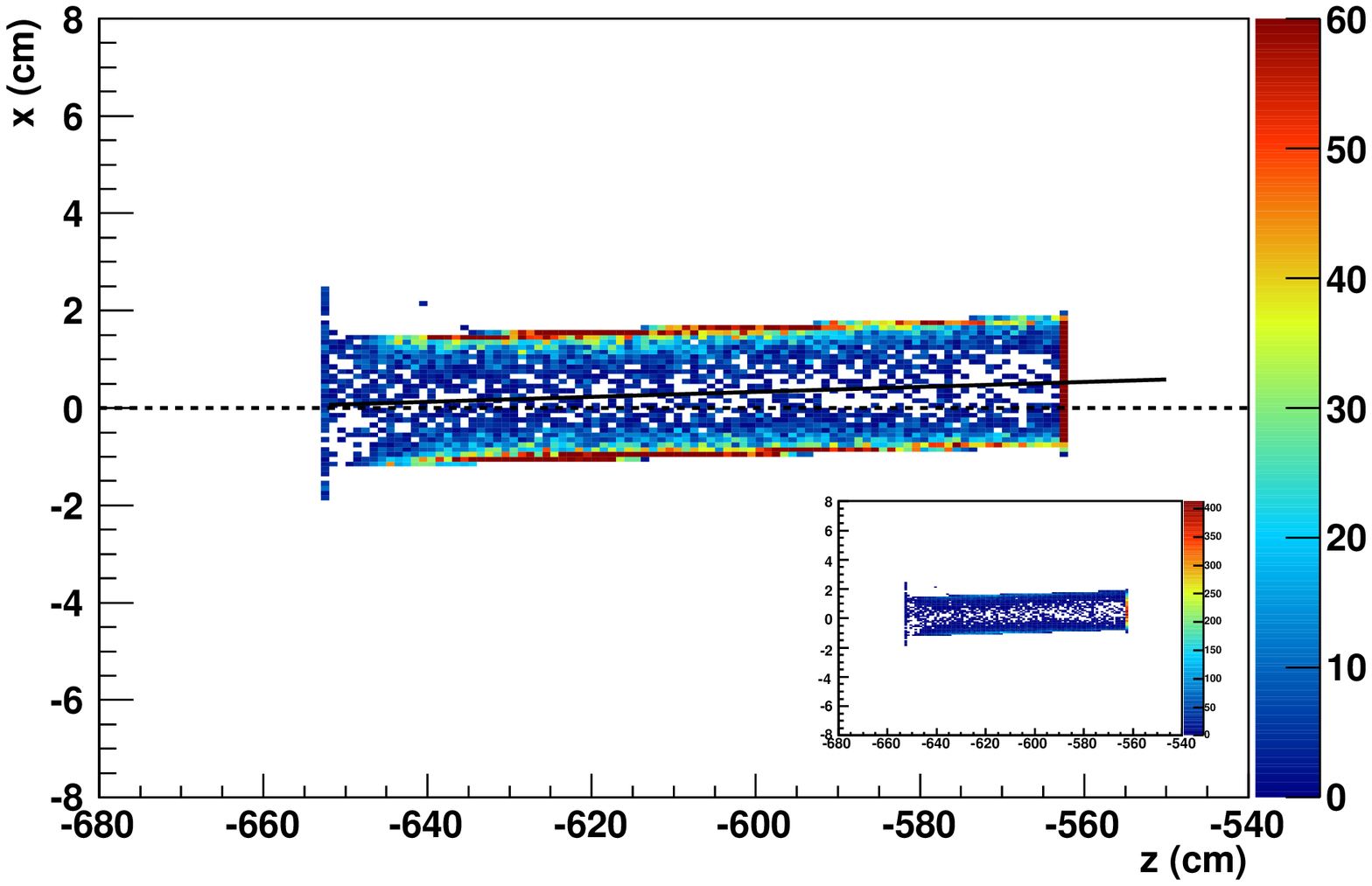}
\includegraphics[width=13pc]{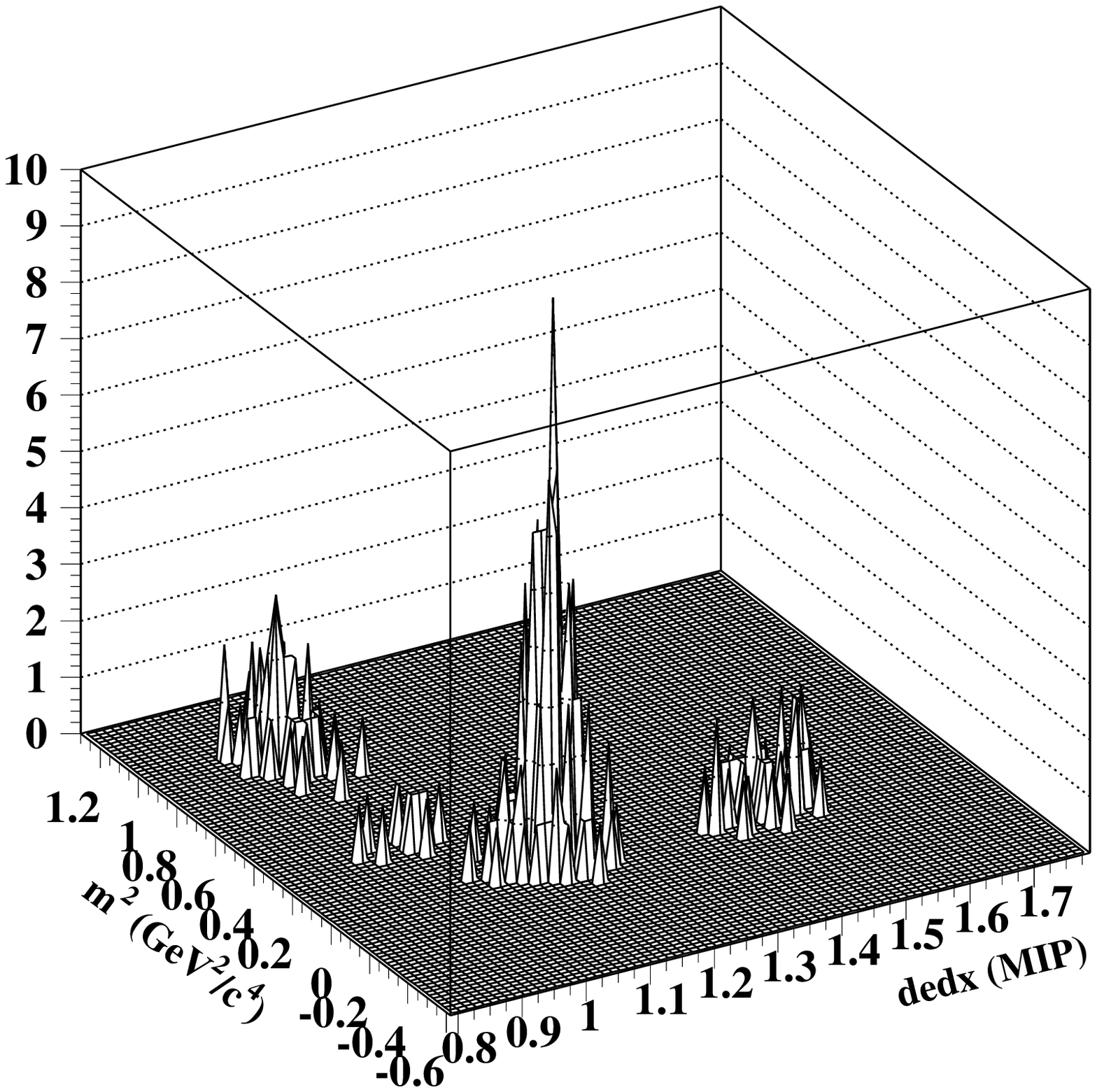}
\hspace{-1pc}
\includegraphics[width=13pc]{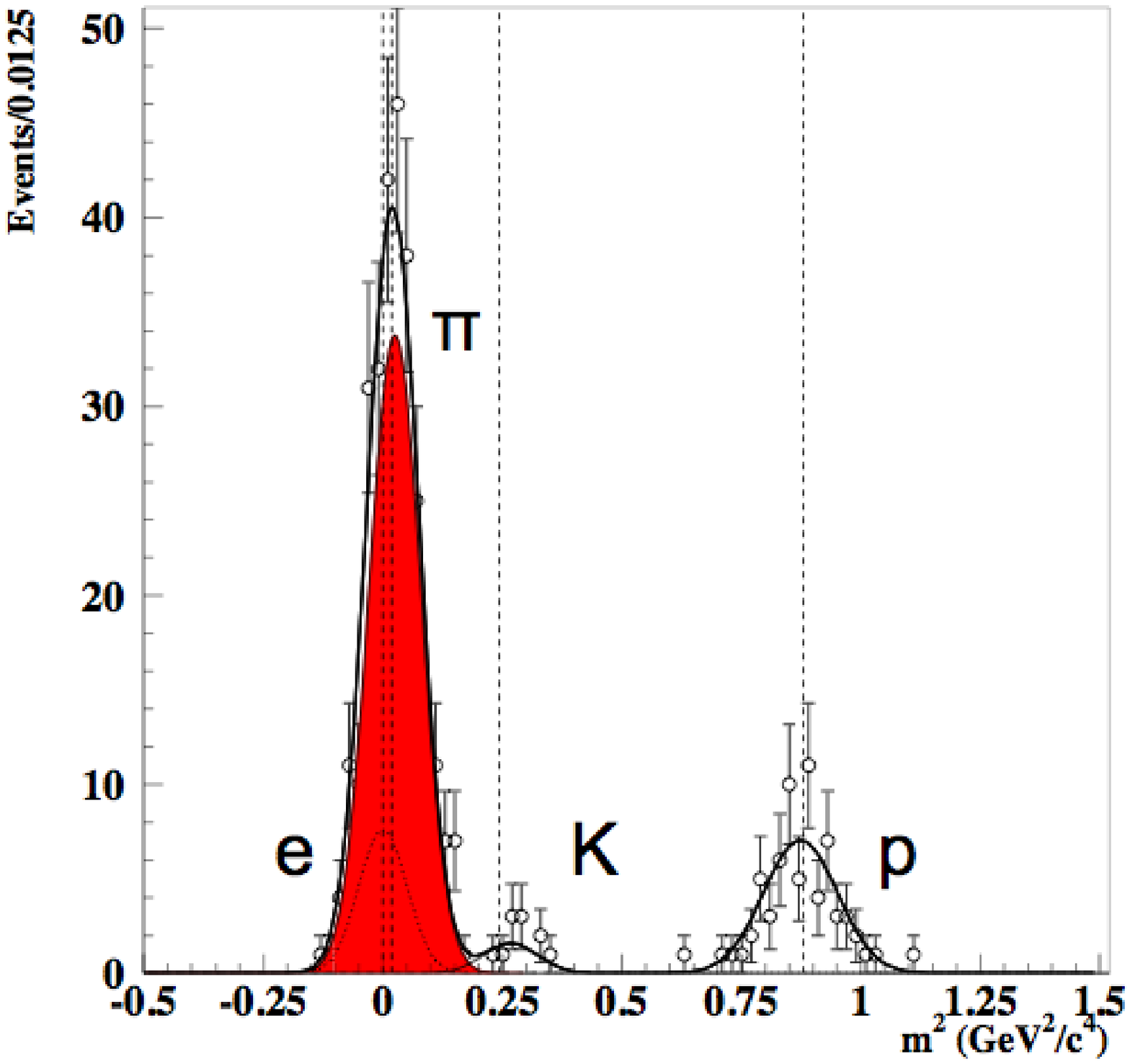}
\caption{\label{reconstruction} Distribution of the point-of-closest-approach on target in
  the x-z projection [left], data distribution in the $\{ToF,dE/dx\}$
  phase space for $40<\theta<100$ mrad and $2.4<p<3.2$ GeV/c [middle],
and projection of the fit result along the ToF axis [right].}
\end{figure}

The particle identification (PID) relies on the energy loss ($dE/dx$)
measurement in the TPCs and the time-of-flight that is used to
compute the particle mass squared. The combination of the two signals
provides a powerful PID over a large momentum range.
Yields of positively charged pions exiting the target are
extracted in bins of $(p,\theta,z)$ using a 2d-likelihood minimization
(see Fig.~\ref{reconstruction}, middle and right panels).
Fig.~\ref{tuning} shows that such yields can
effectively be measured at the surface of the target.

The FLUKA simulation package \cite{FLUKA} used in the T2K beam
simulation was interfaced to the NA61/SHINE simulation chain for which
a PID based on data parametrizations was furthermore implemented.
Data and Monte-Carlo are processed identically with the
same reconstruction chain and PID analysis. Thus, reconstructed yields
of positively charged pions can be compared to
reconstructed Monte-Carlo predictions prior to any corrections necessary to get absolute yields
(e.g. all Monte-Carlo based corrections such as acceptance and decay corrections, reconstruction efficiency,
etc). Such a comparison is depicted in Fig.~\ref{tuning} (top left panel) with the corresponding re-weighting 
factors (top right panel), defined as the ratio of data to Monte-Carlo yields.
Current uncertainties on the re-weighting factors are dominated by
the statistical uncertainty which is typically 10\% due to the very
poor statistics of the 2007 pilot data.  Systematic uncertainties are within 5-10\%, the dominant source being 
PID (1-5\%), followed by target misalignment (3\%) and normalization (1.4\%). Systematics from ToF efficiency, 
beam momentum and target density are all below 3\%.

Comparisons of FLUKA-standalone predictions for the T2K beam simulation to the NA61/SHINE long
target data are also shown as an example in Fig.~\ref{tuning} (bottom panel), for
predictions re-weighted with the NA61/SHINE thin target and long target data.
More details on the NA61/SHINE long target data analysis and re-weighting procedure are
given in \cite{thesis-Abgrall}.

\vspace{1.5pc}
\begin{figure}[h]
\label{tuning}
\vspace{-2pc}
  \centering
\includegraphics[width=14pc]{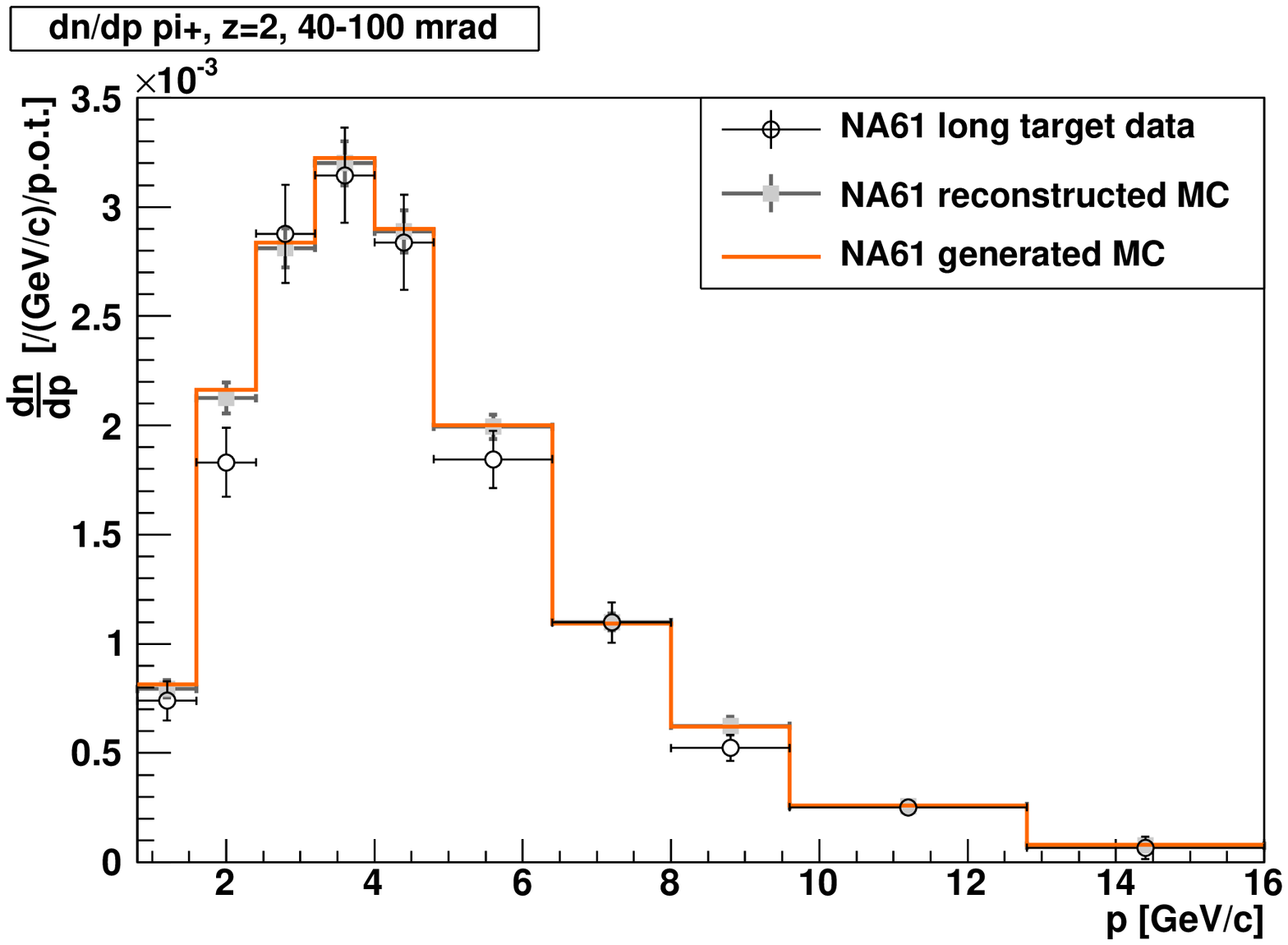}
\includegraphics[width=14pc]{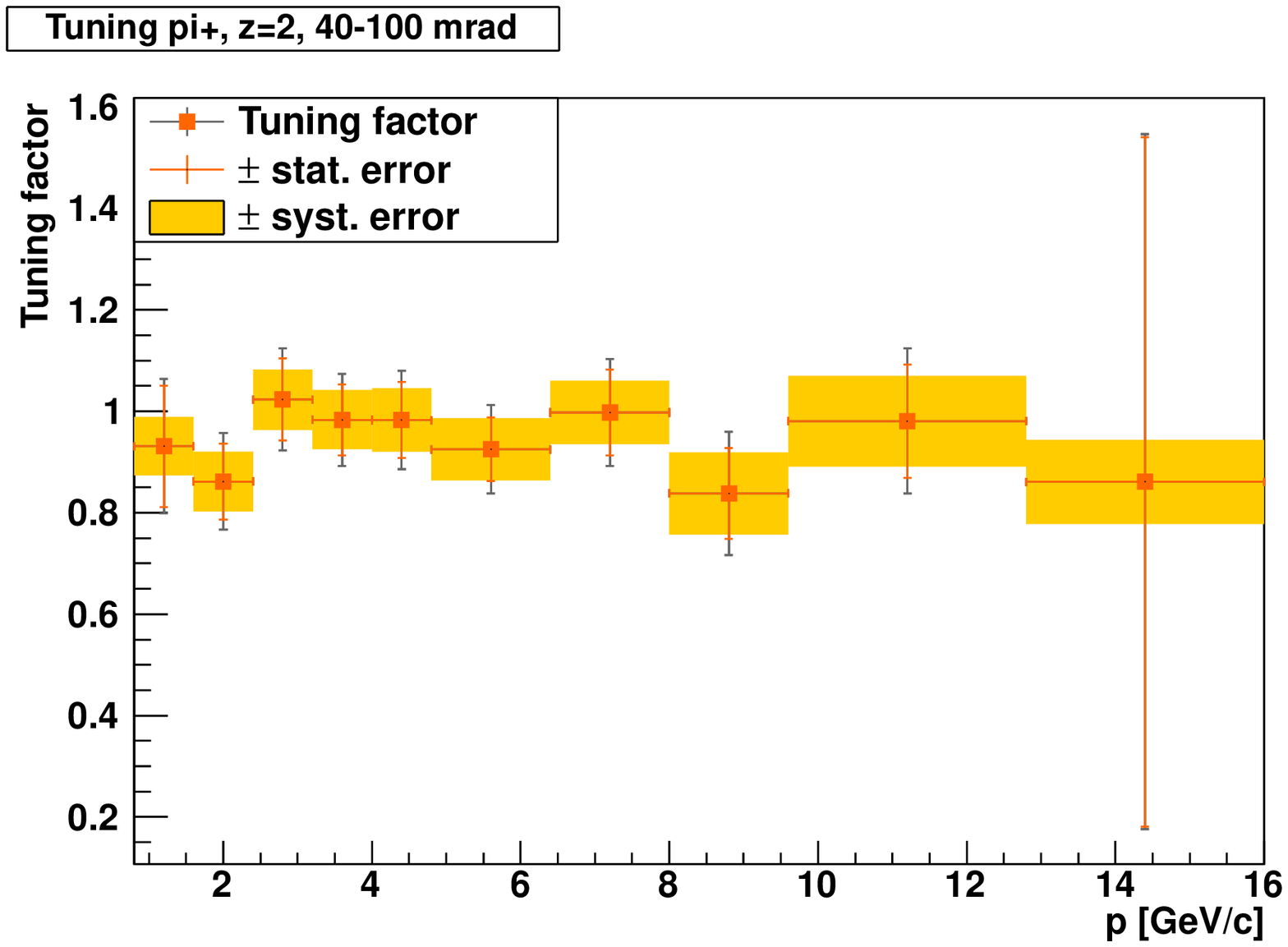}
\includegraphics[width=14pc]{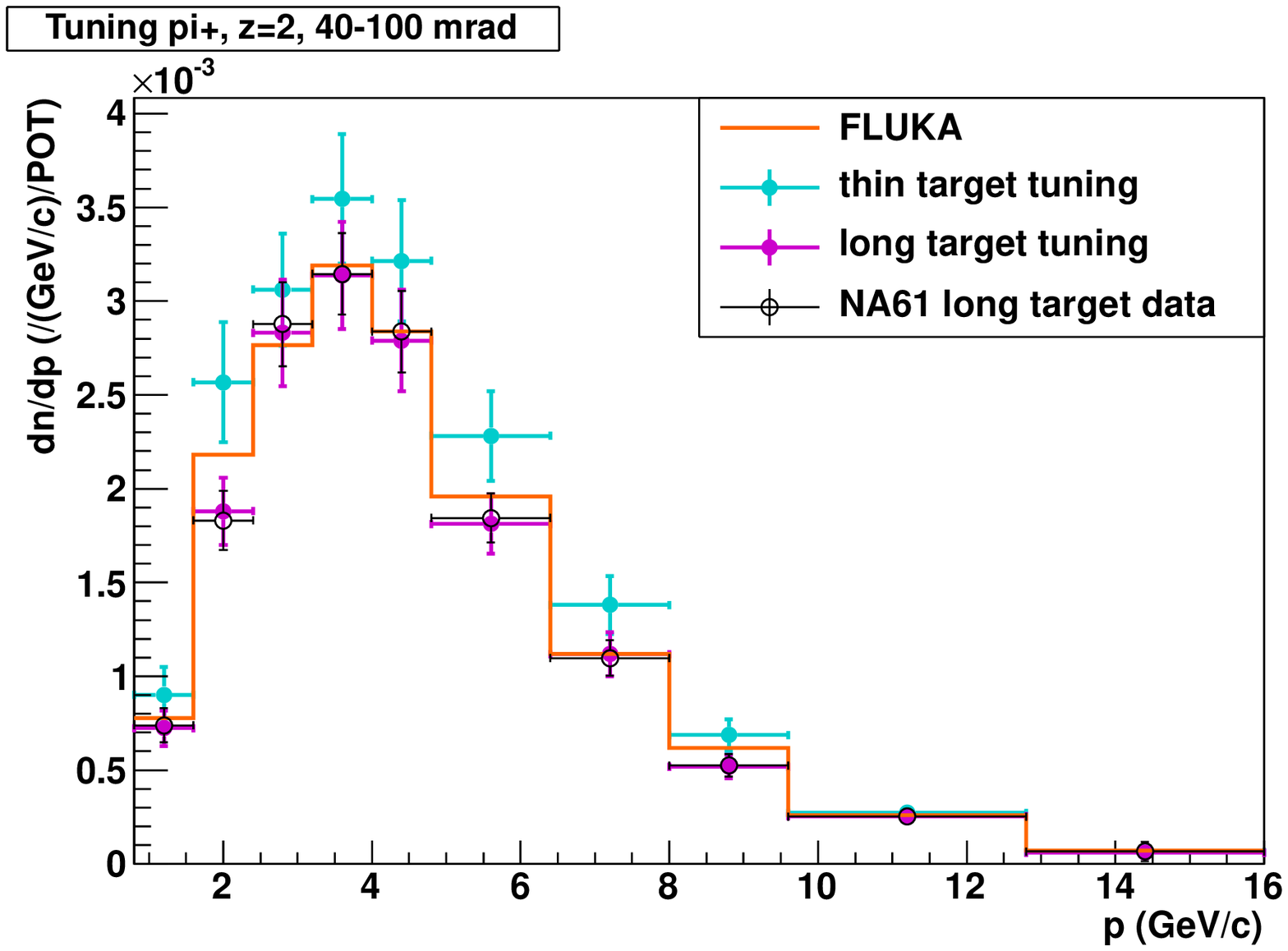}
\includegraphics[width=14pc]{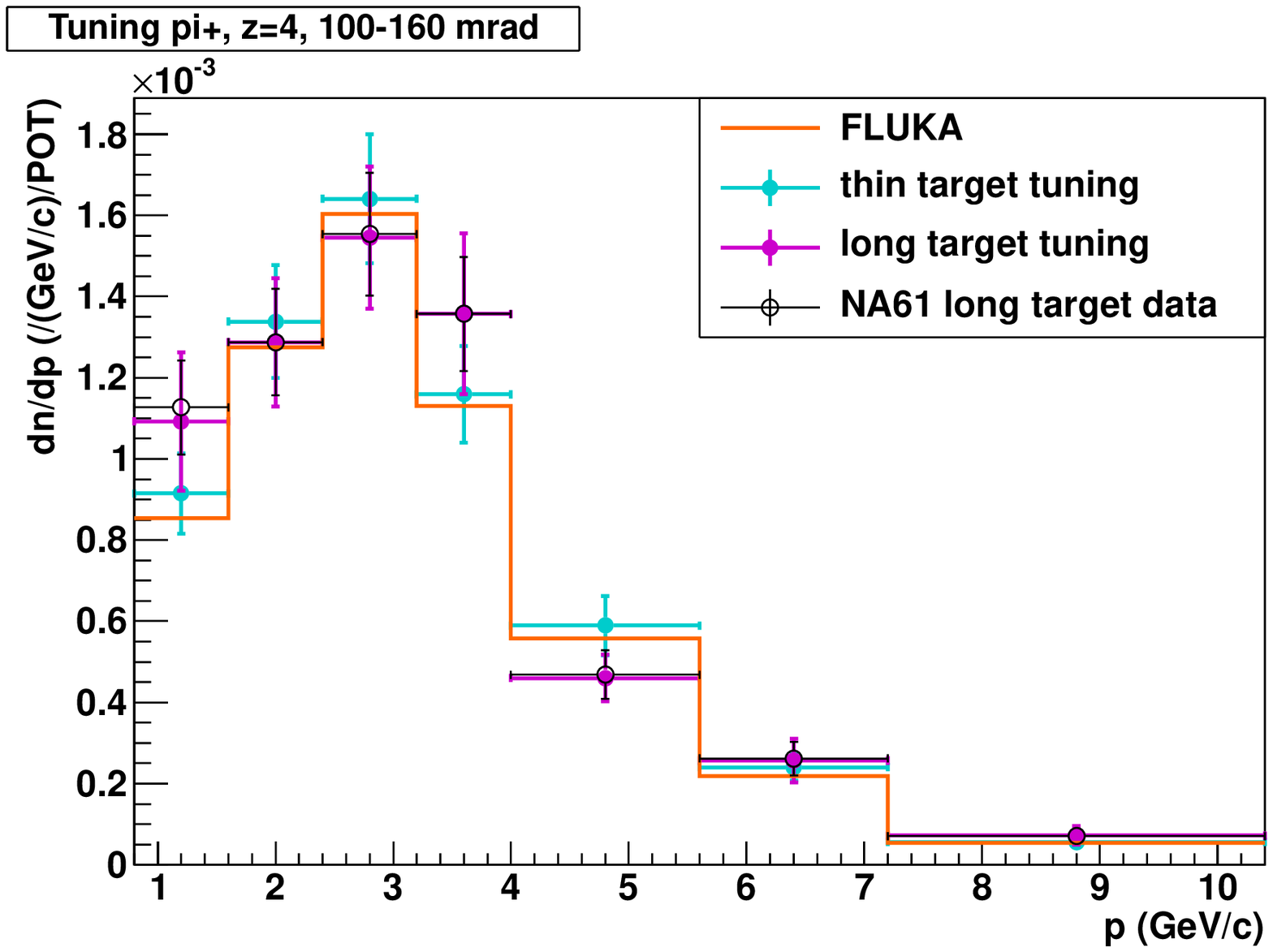}
\caption{\label{tuning} Data and Monte-Carlo yields of positively charged pions normalized to the bin
  size and to the number of protons on target over the second longitudinal
  bin of the target for $40<\theta<100$ mrad [top left], and corresponding re-weighting
  factors [top right]. Comparisons of nominal FLUKA-standalone predictions with thin target and
  long target re-weighting to the NA61/SHINE long target data for different angular intervals [bottom panel].}
\end{figure}

Statistical uncertainties will be reduced with the 2009/2010 long
target data sets, and further improvements are expected (in particular
for the target alignment) to reduce the systematic uncertainties down to
5\% or better. The pilot analysis of the 2007 NA61/SHINE long target
data demonstrates that the T2K neutrino flux predictions can
effectively be re-weighted with long target data. A first preliminary
re-weighting of the $\nu_\mu$ flux prediction at the far detector was
already implemented and compared to the prediction based on the
2007 NA61/SHINE thin target data (\cite{NA61-pion-paper}, \cite{thesis-Abgrall}).

\end{document}